# Ångström-resolved Interfacial Structure in Organic-Inorganic Junctions

Craig P. Schwartz[1*†], Sumana L. Raj[2†], Sasawat Jamnuch[3], Chris J. Hull[2,3], Paolo Miotti[4,5], Royce K. Lam[2,3], Dennis Nordlund[6], Can B. Uzundal[2], Chaitanya Das Pemmaraju[7], Riccardo Mincigrucci[8], Laura Foglia[8], Alberto Simoncig[8], Marcello Coreno[9], Claudio Masciovecchio[8], Luca Giannessi[8,10], Luca Poletto[4], Emiliano Principi[8], Michael Zuerch[2,11,12], Tod A. Pascal[3,13,14], Walter S. Drisdell[1,15*], Richard J. Saykally[1,3*]

1. Chemical Sciences Division, Lawrence Berkeley National Laboratory, Berkeley, California 94720, USA
2. Department of Chemistry, University of California, Berkeley, California 94720, USA
3. ATLAS Materials Science Laboratory, Department of NanoEngineering and Chemical Engineering, University of California, San Diego, La Jolla, California, 92023, USA
4. Institute of Photonics and Nanotechnologies, National Research Council of Italy, via Trasea 7, I-35131 Padova, Italy
5. Department of Information Engineering, University of Padova, via Gradenigo 6/B, I-35131 Padova, Italy
6. Stanford Synchrotron Radiation Lightsource, SLAC National Accelerator Laboratory, Menlo Park, California 94025, USA
7. Theory Institute for Materials and Energy Spectroscopies, SLAC National Accelerator Laboratory, Menlo Park, California 94025, USA
8. Elettra-Sincrotrone Trieste S.C.p.A., Strada Statale 14—km 163.5, 34149 Trieste, Italy
9. ISM-CNR, Istituto di Struttura della Materia, LD2 Unit, Trieste, Italy
10. ENEA, C.R. Frascati, Via E. Fermi 45, 00044 Frascati (Rome), Italy
11. Institute for Optics and Quantum Electronics, Abbe Center of Photonics, University of Jena, Jena, Germany
12. Fritz Haber Institute of the Max Planck Society, Berlin, Germany
13. Materials Science and Engineering, University of California San Diego, La Jolla, California, 92023, USA
14. Sustainable Power and Energy Center, University of California San Diego, La Jolla, California, 92023, USA
15. Joint Center for Artificial Photosynthesis, Lawrence Berkeley National Laboratory, Berkeley, California 94720, USA

Correspondence to: *cpschwartz@lbl.gov (C.P.S.), *wsdrisdell@lbl.gov (W.S.D.), *saykally@berkeley.edu (R.J.S.)

† These authors contributed equally.

**Abstract:** Charge transport processes at interfaces which are governed by complex interfacial electronic structure play a crucial role in catalytic reactions, energy storage, photovoltaics, and many biological processes. Here, the first soft X-ray second harmonic generation (SXR-SHG) interfacial spectrum of a buried interface (boron/Parylene-N) is reported. SXR-SHG shows distinct spectral features that are not observed in X-ray absorption spectra, demonstrating its extraordinary interfacial sensitivity. Comparison to electronic structure calculations indicates a boron-organic separation distance of 1.9±0.1 Å, wherein changes as small as 0.1 Å result in easily detectable SXR-SHG spectral shifts (ca. 100s of meV). As SXR-SHG is inherently ultrafast and sensitive to individual atomic layers, it creates the possibility to study a variety of interfacial processes, e.g. catalysis, with ultrafast time resolution and bond specificity.

Surfaces and interfaces play central roles in a variety of critical biological systems, electronics, batteries, and catalytic systems. Key chemical reactions and physical processes depend explicitly on the electronic structure of the interface and the dynamics across it. Experimentally, surfaces are often studied using a range of spectroscopic and imaging techniques, from grazing incidence X-ray scattering(1, 2), visible and IR second harmonic generation (SHG) and sum frequency generation (SFG)(3, 4), to scanning probe(5, 6) and total internal reflection (7–9) spectroscopies. However, these techniques are limited in the context of *buried* functional interfaces because of interactions with the bulk materials (e.g. absorption or limited escape depth for photoelectrons) that can attenuate or obfuscate the signal from the interface. Moreover, optical nonlinear spectroscopies are well-suited to study vibrational dynamics, however, details of electronic structure specific to interfacial bonds remain unresolved.

Until recently, X-ray nonlinear spectroscopy was precluded by the lack of available light sources with sufficient coherence and flux, but the recent advent of X-ray free electron lasers (FELs) that generate femtosecond pulses with high peak powers and coherence has enabled such experiments(10–12). Soft X-ray SHG (SXR-SHG) offers powerful advantages compared to other surface-specific techniques(13, 14): it has high penetration depth and combines the element specificity of X-ray absorption spectroscopy with the interfacial specificity of second order nonlinear spectroscopies. Linear X-ray absorption spectroscopy (XAS) is a generally useful tool for studying compounds as it is element-specific and sensitive to the chemical and molecular environment of a target atom. Element-specific measurements of core-to-valence transitions can resolve individual contributions to the electronic structure, which is not easily possible in optical spectroscopies that detect valence-to-valence or vibrational transitions. This is especially important for disentangling the contributions from hybridization at interfaces. While X-rays are highly penetrating, the use of different detection methods provide a range of depth sensitivities: transmission measurements are bulk-sensitive, fluorescence detection is sensitive to approximately 1 μm thick slabs based on the penetration of the photons(15), and photoelectron or total electron yield (TEY) detection provides sensitivity of a few nanometers due to the limited escape depth of photoelectrons(16–20). For hard X-ray radiation (> 5 keV; < 0.25 nm), the wavelength is short enough that the electric field is sensitive to inhomogeneities on an atomic scale, and the second harmonic (SH) is generated throughout the material, rendering it bulk-sensitive(21, 22). As a result, none of these techniques can match the interfacial specificity of visible and SXR-SHG, which have sufficiently long wavelengths that the second harmonic signal is primarily generated at interfaces of centrosymmetric media, as shown recently in the first demonstration of SXR-SHG(13). SXR-SHG provides detailed electronic structure information analogous to that probed by X-ray absorption, with specificity for interfaces but no requirement for smooth surfaces. It is therefore ideally suited for filling an important need, viz. studying buried interfaces, but an experimental proof is thus far lacking.

Here, we demonstrate the application of SXR-SHG for probing the buried interface of a boron film with a support layer of Parylene-N, a prototypical organic-inorganic interface. This buried interface cannot be easily studied with visible SHG because visible light will be strongly absorbed by the materials. In the experiment, we compare the SXR-SHG spectra of the boron/vacuum (B/V) and boron/Parylene-N (B/PN) interfaces (Fig. 1), providing the first demonstration of probing the element-resolved electronic structure at a buried interface. TEY XAS measurements, typically considered surface-sensitive, are largely indistinguishable for the two interfaces because the several nanometer probe depth is too large. In contrast, clear differences are observed in the SXR-SHG spectra, allowing a detailed determination of the interfacial bond characteristics. Accompanying detailed first principles' calculations of SXR-SHG spectra for both interfaces permits a detailed interpretation, showing that the observed experimental shift between the two spectra is due to boron interactions with Parylene-N, indicating a strong surface spectral sensitivity to weak interactions, like London dispersion forces.

At the EIS-TIMEX beamline at the free-electron laser (FEL) FERMI(*23*), the SHG signal from the sample was detected using the same apparatus as in our recently reported study(*13*). Nine different soft X-ray photon energies in the range from 184 to 200 eV were used. These fundamental input energies were just below, at, and just above the boron K-edge(*24*). The input intensity ($I_0$) of the FEL was determined from the drain current of an ellipsoidal mirror upstream of the sample. The samples comprised an unsupported 200 nm boron film and a 200 nm boron film with a 100 nm Parylene-N support layer, purchased from Lebow Corporation. TEY X-ray absorption spectra of the two materials were collected at Beamline 8-2 at SSRL with the Parylene-N layer 10 nm to enhance signal and the sample mounted on silicon (*25, 26*).

In our previous X-ray SHG study, it was found that the SHG signals were very sensitive to the quality of the FEL laser pulse(*13*). Therefore, each FEL shot was filtered by the spectrum of the pulse collected before the sample(*27*). The intensity of the generated SHG response is given by the relation

$$I_{SHG} \propto \left|\chi^{(2)}\right|^2 I_0^2$$

where $\left|\chi^{(2)}\right|$ is the nonlinear susceptibility of the interfacial layer of boron atoms. The SXR-SHG intensity was plotted proportionally to $I_0^2$ (assuming constant pulse length, spot size) for each input energy and fit with a linear regression. As can be seen in the equation, the resulting slope is proportional to $\left|\chi^{(2)}\right|^2$. Finally, this slope was plotted as a function of photon energy to generate the nonlinear spectrum of the material properties at the surface or interface.

The measured SXR-SHG spectra of the B/V and B/PN interfaces are shown in Figure 2. Resonance effects can be seen in both spectra, as the SXR-SHG intensity increases when the fundamental energy is at or above the boron K-edge. There is an increase in cross section of B/PN compared to B/V at 188 eV, 191 eV and a decrease at 193.5 eV. Most notably, we observe a substantial increase in the nonlinear response at energies slightly above the B K-absorption edge. We attribute this to dipole-allowed resonant transitions from 1s to unoccupied states with B p-character. The nonlinear response of this spectral region therefore becomes highly sensitive to the electronic valence structure of the interfacial bonds. The second photon absorption process, into a virtual state well above the conduction band to complete the SHG process, is non-resonant and thus less sensitive to the interfacial bonds. Well below and well above the edge, the SXR-SHG spectra of B/V and B/PN are within error of each other.

In contrast to SXR-SHG results, there are no major differences in the linear TEY spectra of the two materials (Fig. 2). The TEY spectra are essentially identical for both samples, indicating that this technique is insufficiently sensitive to the interface to capture the differences seen in the present SXR-SHG spectra. It should be noted that it is possible for some SXR-SHG signal to be generated at the back B/V interface, but this contribution will be smaller than that of the front interface due to absorption of the fundamental by the boron layer. More specifically, and as we show, the SXR-SHG signal is relevant for analysis at energies above the linear absorption edge, and we operate in direct resonance of the 1s to 2p transition (K edge). The attenuation length for X-rays above the edge is approximately 50 nm. Therefore, the FEL beam is largely absorbed in the boron film and the remaining intensity on the rear side is expected to have negligible contribution to the SXR-SHG signal. In contrast, the generated SXR-SHG signal (368 to 392 eV) from the front side is well above the B absorption edge and is therefore only weakly absorbed by the boron slab. Additionally, the back B/V interface is the same for both samples, and so will not affect any qualitative comparisons between the two interfaces.

First principles' electronic structure calculations via perturbation theory within density functional theory(*13, 28, 29*) were employed to simulate the SHG response function. Here, the B/V SHG calculation was performed using two layers of boron icosahedral unit cells. In order to understand the influence of the organic molecule on the electronic structure and resulting SXR-SHG spectra at the interface, we use boron/ethane (B/E) as a proxy for boron/Parylene-N, for computational feasibility, which have a similar lineshape. Given resonant conditions, the second harmonic signal in this energy range is expected to arise primarily from the top boron layer[13], such that the difference between one layer and multiple layers of ethane in the simulation is negligible.

Our electronic structure calculations reveal a red shift of the SXR-SHG spectrum at the boron K-edge for B/E, as compared to B/V (Fig. 3A). This is in general agreement with the experimental spectra for B/PN vs. B/V, where we find a redshift of less than 2 eV of the main SXR-SHG peak at 191 eV. We explored the effect of the interfacial bond length between boron and the organic layer on the simulated spectra in Fig.3B, where we find a monotonic redshift in the 191 eV peak with increasing ethane separation, such that an increase in the separation distance from 0.9Å to the optimal 1.9Å resulted in a 2eV shift, while a further increase to 2.9Å lead to a further 1eV shift. Of note, an increase in the separation distance from 1.9Å to 2.0Å resulted in a 200 meV spectral shift. This predicted high sensitivity of SXR-SHG to interfacial bond length indicates a unique and general technique for elucidating interfacial structure. Our calculations reveal that that ethane induces a shift in the Hartree potential near the interface due to electronic screening, the magnitude of which is strongly dependent on the proximity of the ethane layer to the boron surface. Density of states calculations (Figure S10) indicate that the boron core energy levels lie at lower energy in B/E compared to B/V as a result of this screening. Thus, the experimental spectral differences between B/V and B/PN can be attributed to electronic screening in the interfacial boron atoms, rather than to specific interactions, e.g. orbital hybridization.

In conclusion, linear XAS and spectra of the B/V interface and the B/PN buried interface exhibited no observable difference between the two interfaces, whereas SXR-SHG reveals distinct differences. This is the first time that a buried interface can be resolved with atom-specific sensitivity. More generally, these experiments clearly demonstrate the sensitivity of SXR-SHG to subtle changes in the interfacial electronic structure of the buried interface with sensitivity to a single atomic layer. The results show that SXR-SHG is highly sensitive to interfacial bond lengths and to sub-Ångström bond length changes, resulting in measurable spectral shifts in the hundreds of meV range. Under the assumption of comparability of an ethane/boron interfacial bond with a Parylene-N/boron interfacial bond, it was possible to determine the bond length to be approximately 1.9±0.1 Å. However, the fact that ethane had to be used as a proxy for Parylene-N in simulations of varying distance highlights the importance of developing numerical methods to enable computation of larger systems with high fidelity. While SXR-SHG spectroscopy clearly is a unique and powerful new tool, profound understanding of interfacial electrodynamics will require a carefully orchestrated duet of theory and experiment. Because of the ultrafast nature of the probe and its sensitivity to single atomic layers, the technique has great potential for future studies of dynamics of buried interfaces in electrochemical cells and catalysts. In the near future, SXR-SHG spectroscopy can be used to probe the interfacial electronic structure in a variety of other systems of critical interest, including electronics, batteries and photocatalytic systems that are difficult to study with other methods. Moreover, the newly-revealed high sensitivity to interfacial bond lengths and symmetries will enable unique studies on interfacial strain and its influence on electronic transport properties.

**Acknowledgments:** Soft x-ray SHG measurements were conducted at the EIS-TIMEX beamline at FERMI. The research leading to these results has received funding from the European Community's Seventh Framework Programme (FP7/2007-2013) under grant agreement nº 312284. TEY X-ray absorption spectra were collected at beamline 8-2 at the Stanford Synchrotron Radiation Lightsource, SLAC National Accelerator Laboratory, which is supported by the US DOE, Office of Science, Office of Basic Energy Sciences under Contract DE-AC02-76SF00515. C. J. H. and S. L. R. were supported by the U.S. Army Research Laboratory (ARL) and the U.S. Army Research Office (ARO) under Contracts/Grants No. W911NF-13-1-0483 and No. W911NF-17-1-0163. R. K. L. and R. J. S. were supported by the Office of Science, Office of Basic Energy Sciences, Division of Chemical Sciences, Geosciences, and Biosciences of the U.S. Department of Energy at the Lawrence Berkeley National Laboratory (LBNL) under Contract No. DE-AC02- 05CH11231. W.S.D. was supported by the Joint Center for Artificial Photosynthesis, a DOE Energy Innovation Hub, supported through the Office of Science of the US DOE under Award DE-SC0004993. S. L. R. received a National Science Foundation Graduate Research Fellowship under Grant No. DGE 1106400. Any opinions, findings, and conclusions or recommendations expressed in this material are those of the author(s) and do not necessarily reflect the views of the National Science Foundation. M.Z. acknowledges support by the Max Planck Society (Max Planck Research Group) and the Federal Ministry of Education and Research (BMBF) under "Make our Planet Great Again – German Research Initiative" (Grant No. 57427209 "QUESTforENERGY") implemented by DAAD. P.M. and L. P. were supported by the project Single-Shot X-Ray Emission-Spectroscopy experiments funded by the Italian Ministry for Education and Research as an in-kind project for the EuroFEL consortium. Simulations were performed as part of a user project with S. J., T. A. P., and C. D. P. at The Molecular Foundry (TMF), LBNL. Theoretical simulations of nonlinear susceptibility by C. D. P. were carried out within TIMES at SLAC National Accelerator Laboratory supported by the U.S. Department of Energy, Office of Basic Energy Sciences, Division of Materials Sciences and Engineering, under Contract No. DE-AC02-76SF00515. Portions of the SXR-SHG spectra calculations used resources of the National Energy Research Scientific Computing Center, which is supported by the Office of Science of the U.S. Department of Energy under Contract No. DE-AC02-05CH11231. This work also used the Extreme Science and Engineering Discovery Environment (XSEDE), which is supported by National Science Foundation grant number ACI-1548562. Material support was provided by the Office of Science, Office of Basic Energy Sciences, Division of Chemical Sciences, Geosciences, and Biosciences of the U.S. Department of Energy at LBNL under the contract listed above. Travel support was provided by ARL and ARO under the Contracts/Grants listed above.

**Author contributions** C.P.S., W.S.D., S.L.R., R.K.L., R.J.S. and M.C. planned the research. C.P.S., S.L.R., C.J.H., R.K.L, P.M., R.M., L.F., A. S., M.C., C.M., L.G., L.P., E.P., and W.S.D acquired the FEL data. D.N. acquired the synchrotron data. S.J., C.D.P., and T.A.P planned and acquired the theoretical simulations. C.P.S., S.L.R., S.J., C.D.P., E.P., M. Z., T.A.P., W.S.D., and R.J.S analyzed and interpreted the results. C.P.S., S.L.R., R.J.S., S.J., C.B.U., T.A.P., M.Z. and W.S.D. wrote the manuscript. R.J.S. supervised the research. All authors contributed to the discussion of results.

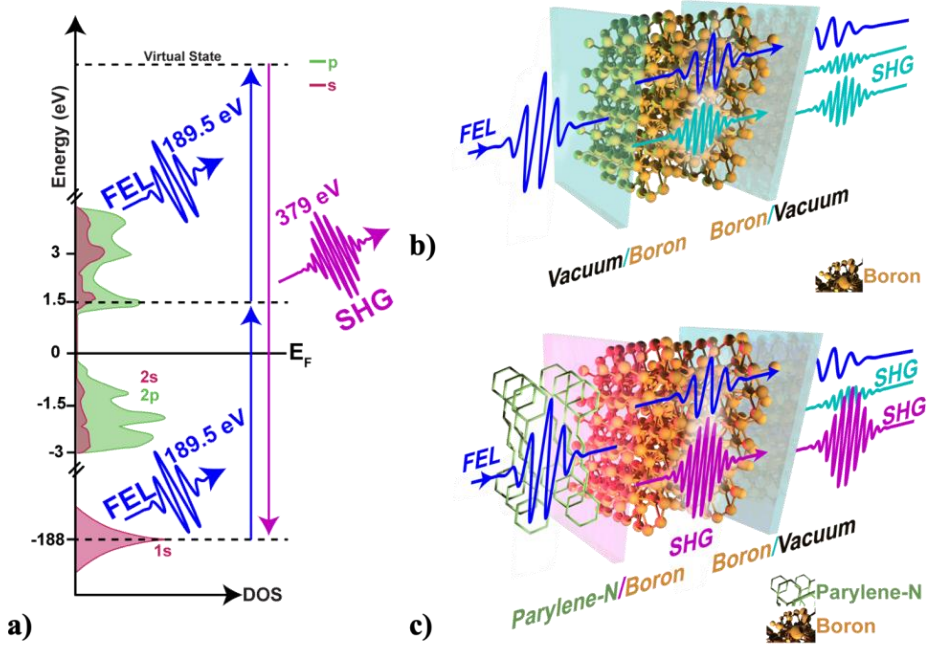

**Fig. 1.** Schematic of SHG from interfaces. In the energy level diagram (a), the density of states of boron (s-type red, p-type green) is resonantly pumped with an FEL pulse (blue). Due to selection rules, only the p-type states are probed. Two photons at this energy combine in the material and a second harmonic photon at twice the energy (purple) is emitted. The input energy is shown at 189.5 eV, generating a photon of 379 eV. During the experiment, the FEL energy was scanned from 184-196 eV. Two different interfaces were studied here, the (b) boron/vacuum interface and a (c) Parylene-N/boron interface. The back boron/vacuum interface also generates some SHG signal (shown in light blue), but it will be less intense due to attenuation of the FEL pulse from transmission through the sample.

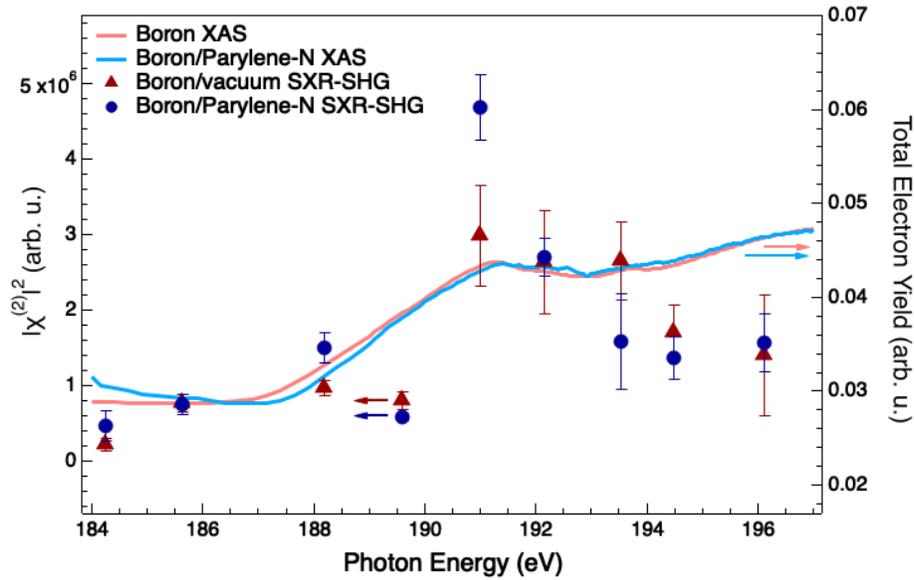

**Fig. 2.** Second harmonic generation spectra of the boron/vacuum and boron/Parylene-N interfaces. The SHG spectrum of the B/V (dark red triangle) and B/PN (dark blue circle) interfaces, shown along with the linear X-ray absorption of the boron film (light red) and boron Parylene-N multilayer film (light blue). $|\chi^{(2)}|$, determined from the linear regression slope of the SHG signal vs. $I_0^2$. $|\chi^{(2)}|$, is significantly higher at the boron K-edge for B/PN than for B/V.

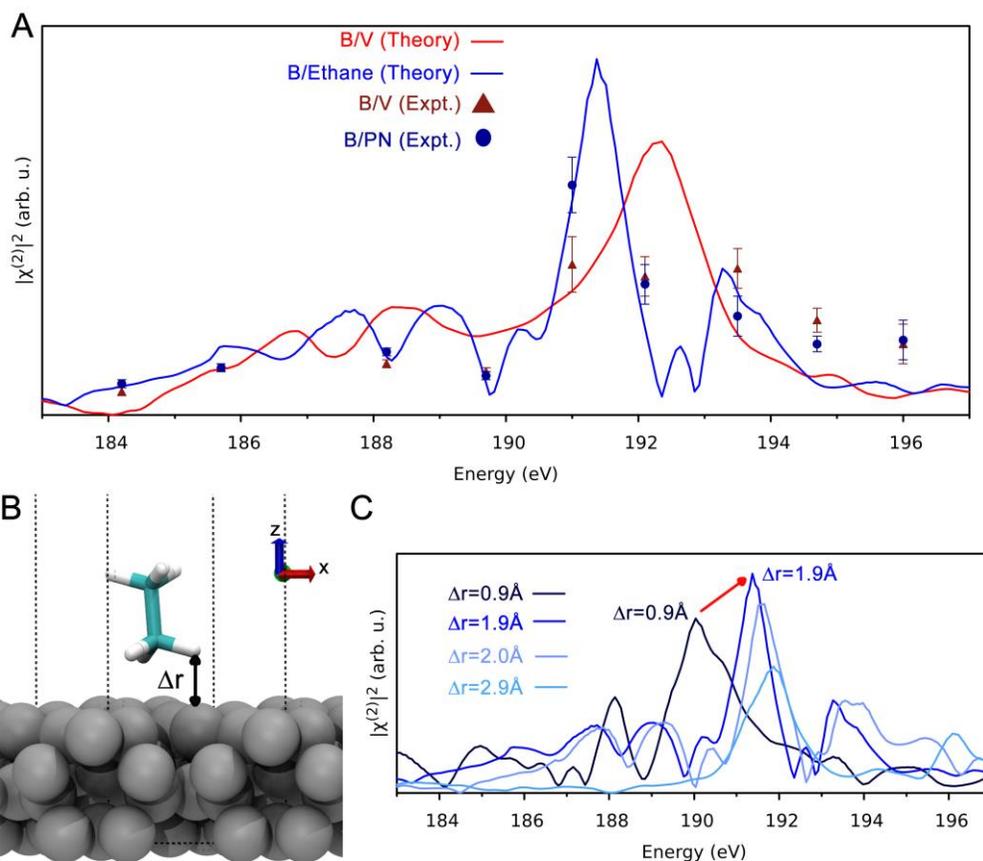

**Figure 3:** A) Comparison of theoretical calculations of SXR-SHG boron/vacuum (red line) spectra and boron/ethane spectra (blue) at its equilibrium distance 1.9 Å with the experimental SXR-SHG measurements of B/V (dark red triangles) and B/PN (dark blue circles). The calculation captures the enhancement of and the blue shift in the 191 eV feature in B/PN compared to B/V. B) Schematic of boron/ethane computational unit cell. We vary the Δr distance between the hydrogen atom and the surface of boron. C) Comparison of theoretical SXR-SHG as a function of distance between boron and ethane, ranging from 0.9 Å (dark blue), 1.9 Å, 2.0 Å to 2.9 Å (light blue), with the color growing increasingly light with increasing distance. As distance between the boron and ethane is increased, the primary peak at 191 eV red-shifts in energy (indicated by the red arrow).

# Supplementary Materials for

## Ångström-resolved Interfacial Structure in Organic-Inorganic Junctions

**Materials and Methods**

**Experimental Methods:**

At the EIS-TIMEX endstation at FERMI, in order to access input photon energies from 184 to 195 eV, the 13$^{th}$ harmonic of FEL-1 and the 3$^{rd}$ harmonic of FEL-2 were used. The energy was tuned within this range by using the OPA to adjust the seed laser photon energies. Photon energies from 195 to 200 eV were accessed using the 10$^{th}$ harmonic of FEL-1 and the 4$^{th}$ harmonic of FEL-2. The FEL was focused to ca. 10 μm diameter spot. In order to examine the SHG-SXR intensity as a function of input intensity, the latter was varied using different metallic filters before the sample (100 nm of Zr, 100 nm of Pd, and 200 nm of Pd), in addition to inherent pulse-to-pulse energy fluctuations of the FEL. An iris was also used upstream of the sample to minimize the off-axis light from the FEL incident on the sample and detector. After the sample, the fundamental intensity was attenuated using 200 nm of aluminum before passing through a slit. The filter transmissions were calculated using the fundamental signal detected by the CCD when there was no sample in place. The exact photon energies of the FEL were also calculated using the position of the fundamental signal on the CCD. Both the fundamental and SHG signals were detected simultaneously using a soft X-ray spectrometer (2400 groove/mm grating) and CCD (Princeton Instruments PIXIS-XO 400B)(S*1*).

Standard transmission values for aluminum, boron, and Parylene-N were used to account for the effects of the aluminum filter after the sample, and transmission through the sample itself(S*2*). Only pulses with a single peak in the spectrum were used. The data were also filtered based on the full width at half maximum (FWHM) of the fundamental signal on the CCD to ensure that multiple overlapping modes were not present. Data taken with no sample in the X-ray beam were used to subtract the unwanted SHG signal generated in the FEL from the SHG signal generated at the sample.

**Theoretical Methods:**

First-principles density functional theory (DFT) (S*3*, S*4*) calculations were carried out using ***exciting***(S*5*) full potential all-electron software package which implements linearized augmented planewave methods. For simulations of B/V, two and three layers of icosahedral boron supercells were set up for SHG response calculations. The SHG response is nearly identical (Figure S8). A vacuum of 10 Å was added along the z-direction (perpendicular to the boron surface) to resample a boron-vacuum interface. The B/PN simulation was set up with one layer of icosahedral born and a single Parylene-N monomer (p-xylene). The B/E simulation was setup with two or three layers of ethane. Initially, an UFF forcefield (S*6*) was used to optimize the structure. Next CP2K(S*7*) was used to optimize the structure again to provide a more accurate geometry. For ground state simulation in ***exciting***, the parameter rgkmax was set to 7. This parameter represents the product between $R_{MT}$, the minimum muffin tin radius, and the maximum length of the **G+k** vector of the basis set. The Brillouin zone was sampled using 20x20x1 Γ-point centered k-point grid. The Perdew-Zunger Local Density Approximation (LDA-PW)(S*8*) was used to model the exchange correlation effect within DFT. Core-hole 1s orbitals of boron were included in the valance self-consistent field for Kohn-Sham eigenvalues and eigenvectors to self-consistently update in the simulations. The theoretical calculation of SHG followed the framework of established by Lam *et al*(S*9*). All spectral simulations were performed on geometry-optimized structures.

To calculate the SH response at B K-edge, we employed the formalism from Sharma(S*10*) implemented within ***exciting***. The approach calculates the second-order susceptibility $\chi^{(2)}(2\omega, \omega, \omega)$ from the resonant energies at ω and 2ω excitation. We employed the independent particle approximation in which the excitation energy is given by the difference between their respective Kohn-Sham eigenvalues. The linear absorption spectra were calculated within the same formalism and a rigid 26 eV blue shift was applied to the spectra to match with the experimental results. The boron K edge is situated ~180 eV below the Fermi level, and so resonant excitation occurs when ω is at 188 eV (6.89 nm). To capture the contribution of B 1s excitation near 2ω, 2000 empty Kohn-Sham eigenstates whose energies extend up to ~225 eV above Fermi level were included in the calculation. The electronic density of state (DOS) hundreds of eV above the Fermi level exhibits slow convergence which leads to noisy response functions and impractical calculation of $\chi^{(2)}$ at high energies. Therefore, we added a heuristic imaginary part that is dependent on energy to account for lifetime broadening. Hence the energy expression becomes:

$$E_m \to E_m + i\Gamma(E_m) \text{ where } \Gamma(E) = \frac{1}{\lambda}\sqrt{E - E_F}$$

Here λ is the inelastic free electron path which is ~8 Å in boron(S*11*). Lastly, the background signal from the valance electrons, which approximately decays as $\sim \frac{1}{Energy}$, was subtracted from output signal. The real and imaginary parts of the output signals were fit with the analytical equation:

$$S = \frac{A_1}{(Energy - A_2)^{A_3}} \text{ with } A_3 \sim 1$$

After background subtraction, the real and imaginary $\chi^{(2)}$ responses were obtained.

In general, SHG requires the breaking of inversion symmetry on the length scale of the incoming radiation(S*12*). As bulk boron possesses inversion symmetry, no second harmonic response should be observed. However, at the surface, the inversion symmetry is broken leading to SHG responses from interfaces. In simulations, the supercell slab system reacts spatially to the homogenous electric field at both the surfaces as the system always exhibits inversion symmetry due to its few-layer size. Consequently, this leads to overall attenuation of the SHG signal. Thus, in order to generate a more accurate SHG response from the simulation, only one surface is selected to actively contribute to the SHG. This is achieved by considering only the B 1s electrons of only the first (top) layer of the two icosahedron boron cells of the slab in the self-consistent field calculation. For the bottom layer, the B 1s electrons are not included hence the inversion symmetry is artificially broken and the interfacial SHG response is obtained. In the multiple component systems, the inversion symmetry does not exist in the supercell slab and the SHG response can be directly calculated without further manipulation. The energy axis of these simulations was calibrated using the same energy shift needed to align the linear absorption spectra with our experiments.

**Supplementary Text**
**Experimental**
A representative CCD image is shown in Figure S1 for the data taken with an input energy of 190 eV and the boron/Parylene-N (B/PN) sample. The total intensity of each signal is determined by summing the intensities of all the pixels in the region of interest (ROI). Baseline corrections were done by subtracting the intensity of a ROI adjacent to that of the signal from the signal intensity. A CCD image was collected for each laser pulse, along with data from PADReS, the laser pulse diagnostic system upstream of the sample. The drain current from a mirror upstream of the sample was also recorded as it is proportional to the pulse intensity.

A sample dataset for a single photon energy is shown in Figure S2. The data were first filtered based on the pulse spectrum upstream of the sample. If the spectrum was not sufficiently gaussian and contains side peaks that are more than 20% of the height of the main peak, the laser pulse is excluded from further analysis. Similarly, the distribution of full width at half maximum (FWHM) of the fundamental signal on the CCD is examined and any outliers of the FWHM box plot are filtered. The data is then binned (Fig. S3), with bins of 400 μJ used at lower intensities and bins of 300 μJ used at higher intensities. The distribution of signals within each bin is also calculated and outliers that lie beyond the outer fence of the box plot are filtered. The binned data is fit with a power series and the fit of the fundamental data taken with no sample is used to calibrate the input pulse energy axis. The SXR-SHG signal from the FEL (i.e., the SHG signal seen in the no sample data sets) is subtracted from the SXR-SHG signal of the sample. The signal is now corrected for the transmission of the sample and the detector efficiency. The SXR-SHG signal is then plot against the input energy squared (Fig. S4). The slope of the linear fit of this data is proportional to $|\chi^{(2)}|^2$. To confirm that the dependence is truly quadratic, as should be the case for SHG, a plot of log(signal) vs. log(input energy) can be fit to a line (Fig. S5). A quadratic function will result in a log-log plot with a slope of 2. As can be seen in Figure S6, the datasets all have slopes close to 2 indicating SXR-SHG.

XAS was taken at Beamline 8-2 of the Stanford Synchrotron Radiation Lightsource by measuring the drain current. To maximize XAS signal, the sample was modified slightly from that used in SXR-SHG. It consisted of 10 nm of Parylene-N on 100 nm of boron backed with doped silicon in contrast to free-standing films used in SXR-SHG. Attempts to measure the Auger electron yield through a cylindrical mirror analyzer resulted in inadequate signal to noise.

**Theoretical**
*Investigation of origin of difference in SH response*
To investigate the shift to SHG, an ethane molecule is used to replace the parylene-N monomer. The methyl group closest to boron surface is kept exactly the same with the boron/parylene case. We show that the SXR-SHG spectra of B/E have similar lineshapes and absorption energies to those of B/PN (Figure S7). Therefore, the close proximity interaction between the two systems remains the same. The SHG response shows very similar result which indicates that the response is surface sensitive to the environment (Figure S8). The ethane orientation was chosen parallel to the surface to minimize the size of the unit cell, and 2 layers of boron were used, showing minor differences from 3 layer calculations

*Dipole correction for surface calculation:*
The supercell approach employed in DFT calculations of asymmetric slabs can often lead to a net surface dipole density and artificial electrical fields(S*13*). To ensure that there was no dipole effect on the calculation, the electrostatic potential of the simulation cell was calculated. As can be seen in Figure S9, the electrostatic potential in the vacuum region is flat. Hence, the vacuum energy level can be established and no further dipole correction to the system is required. The electrostatic potential due to the presence of a dipole in the vertical direction was also calculated, but the result shows that the effect is negligible (see supporting information Figure S9), and so a dipole correction is not required.For the less computationally expensive B/V simulation, no significant difference is found between the SHG signal from two layers and three layers of boron.

*Boron 1s core electronic states theoretical calculation:*
The transition of 1s core electrons to the conduction band determines the energy required for X-ray absorption. Therefore, in order to verify the red shift observed at the B/PN and B/E interface, the density of states of B 1s electrons are calculated in Figure S10. The simulation is performed using *exciting* code at the independent particle approximation. The parameter ngrdos, which is the effective k-point mesh size, is set to 1000 and three point average smearing is used. The energy range in Hartree is -1 to 1 for the density of states around the fermi level (0 eV) and -6.4 to -6.25 for the 1s core. The calculation of DOS of 1s boron electron shows that there is an overall red shift to the core level electron in B/PN system, which is consistent with the calculated changes in the SHG spectra. Additionally, ground state calculations show that there is less charge in boron muffin-tin radius, which indicates that the presence of a coating weakens the surface boron bonds.

To further verify the origin of signal amplification, we calculated the projected DOS decomposed into 2s and 2p orbital contributions of the interfacial and bulk boron atoms (Fig. S11). We found that there is increased s-orbital character near the conduction band in the B/PN and B/E systems, which have better overlap with the excited 1s core electrons and therefore leads to an increase in the SHG intensity. Moreover, our calculations find that there are more of these s-like state at the Fermi level (0 eV) for the B/PN and B/E interfaces. Thus, screening due to the presence of an organic causes a redshift to the spectrum while the increase in the number of electronic states at the Fermi level leads to the amplification in SXR-SHG.

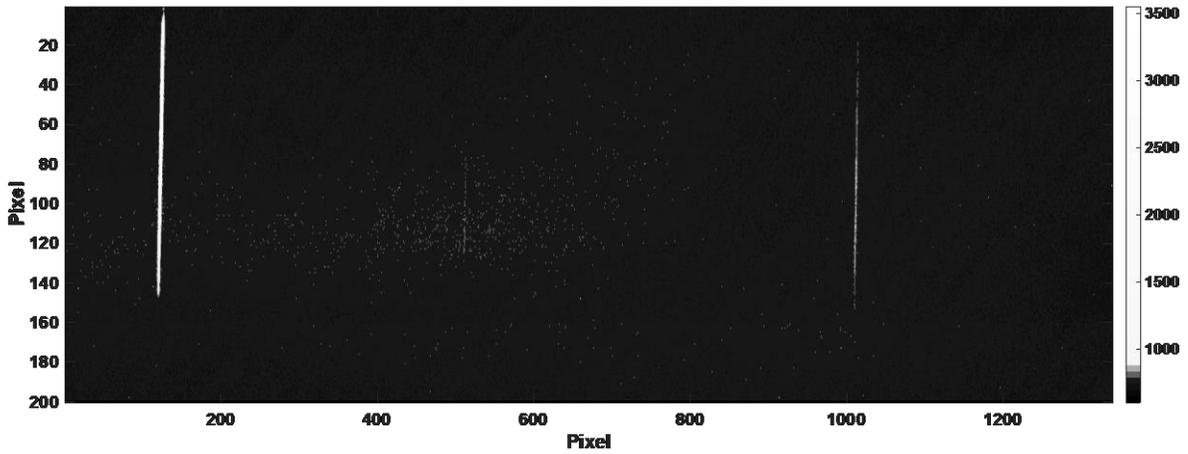

**Figure S1:** Representative CCD Image. Average CCD image of the fundamental and SHG from a single laser shot. The fundamental was at 190 eV and is seen on the CCD at pixels 108 to 133. The SHG from both the sample and FEL source is seen at pixels 1002 to 1012 which corresponds to a photon energy of 386 eV. Baseline correction for the fundamental was performed using adjacent pixels (133 to 158) and the same was done for the SHG using pixels 1022 to 1042.

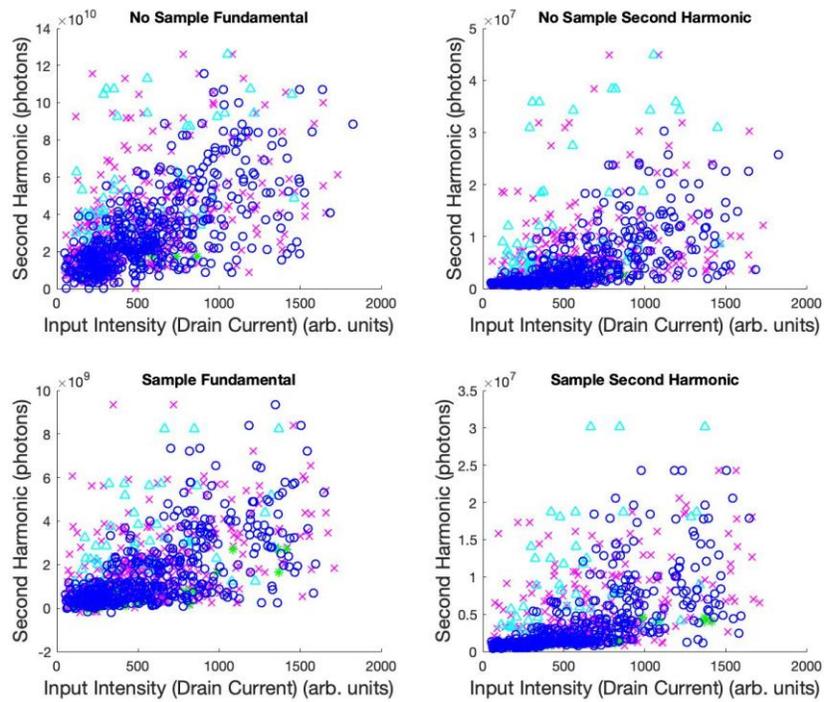

**Figure S2:** Plot of all laser shots vs. input drain current. Data taken at an input photon energy of 191 eV is shown in this plot. The total fundamental signal intensity (a) and total SHG signal intensity (b) with no sample are shown along with the fundamental (c) and SHG (d) signal intensities with the B/PN sample in place. Points were filtered and removed based on the pulse spectrum (x), fundamental signal FWHM (∗), and bin outliers (△). Only the remaining points (o) are used in further calculations.

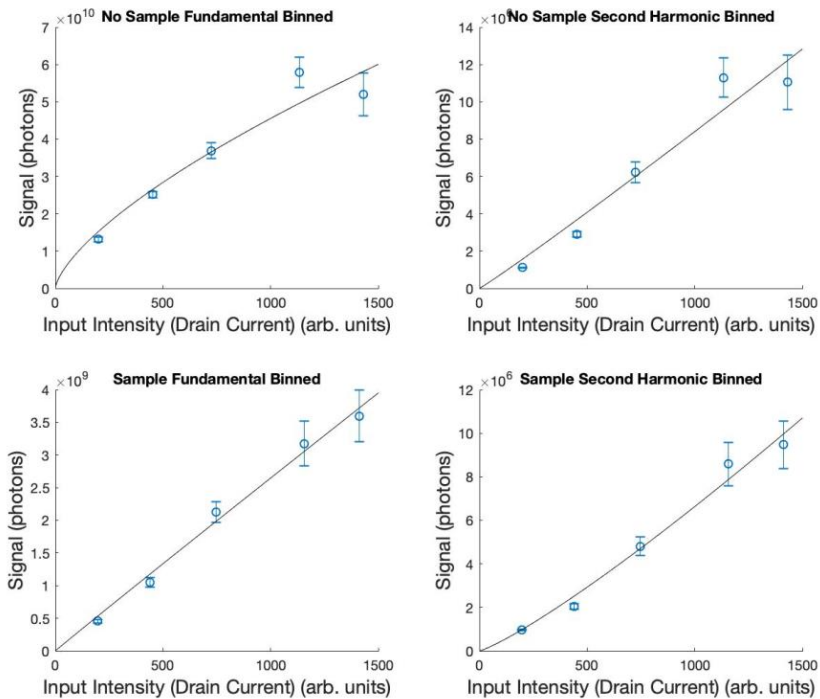

**Figure S3:** Binned and filtered signal vs. input drain current. The binned and filtered signals for the fundamental (a) and SHG (b) without a sample, and the fundamental (c) and SHG (d) with the B/PN sample are shown for an input energy of 191 eV.

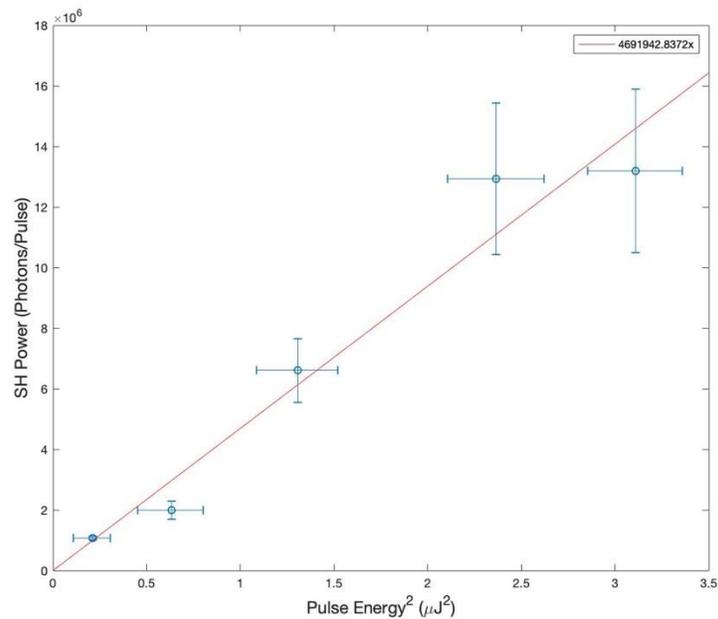

**Figure S4:** SHG generated by the sample vs. pulse energy squared. The SHG signal generated by the B/PN sample with an input photon energy of 191 eV is shown. The slope of the linear fit (4.6 x 10$^6$) is proportional to $|\chi^{(2)}|$.

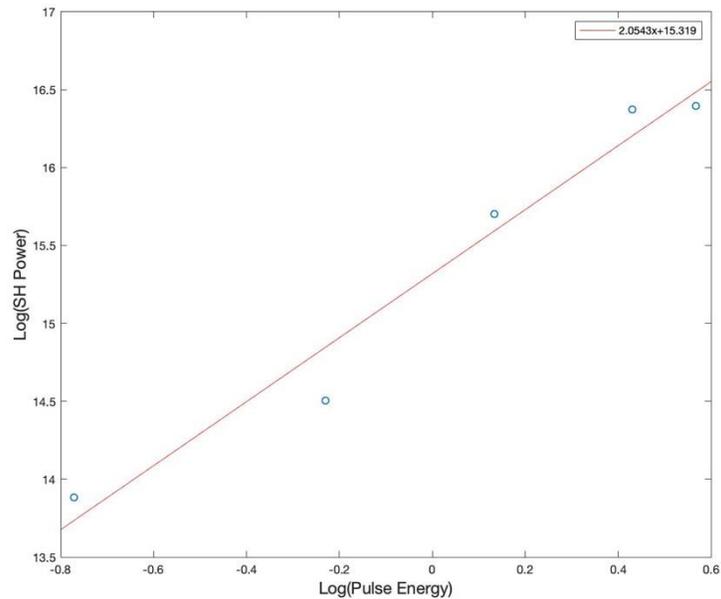

**Figure S5:** Representative log-log plot. The plot of log(SHG signal) vs. log(Pulse Energy) is fit with a line to confirm that the dependence is indeed quadratic. For a quadratic function the slope will be 2. Here, for the data taken at an input energy of 191 eV and with the B/PN sample, the slope is 2.0.

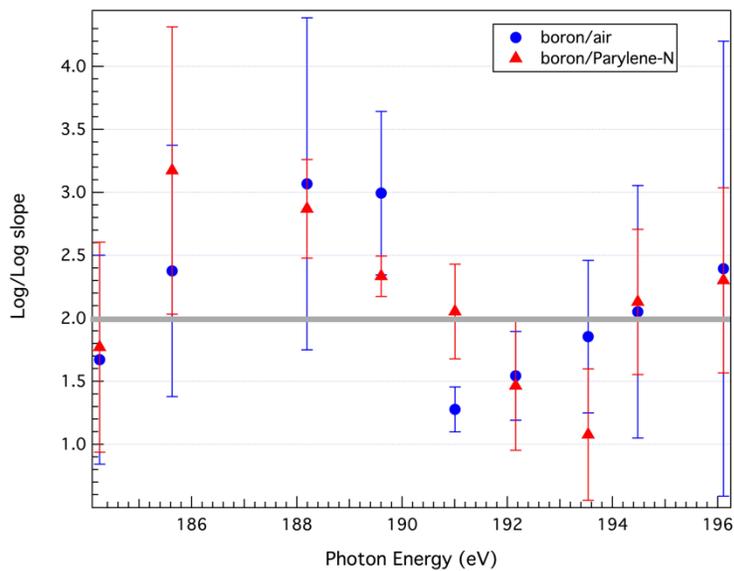

**Figure S6:** Log-log slopes for all datasets, B/V (blue) and B/PN (red). The log-log slopes are shown with their standard errors for all wavelengths and both samples, verifying that the signal we are seeing is in fact SXR-SHG. For a quadratic function, the log-log slope slope will be 2.

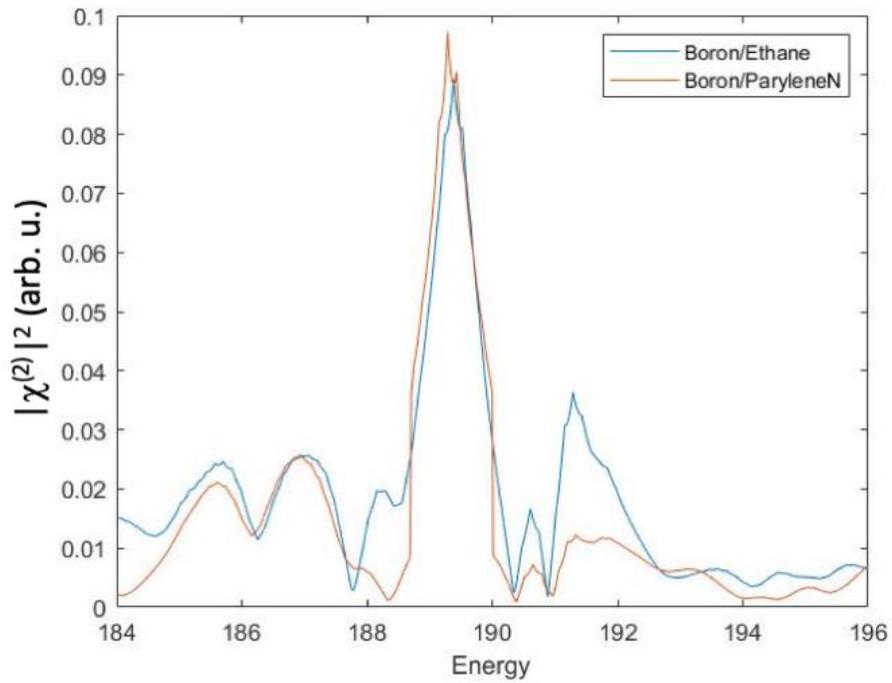

**Figure S7:** Calculated SXR-SHG spectral comparison between B/PN (blue) and B/E (orange) interface.

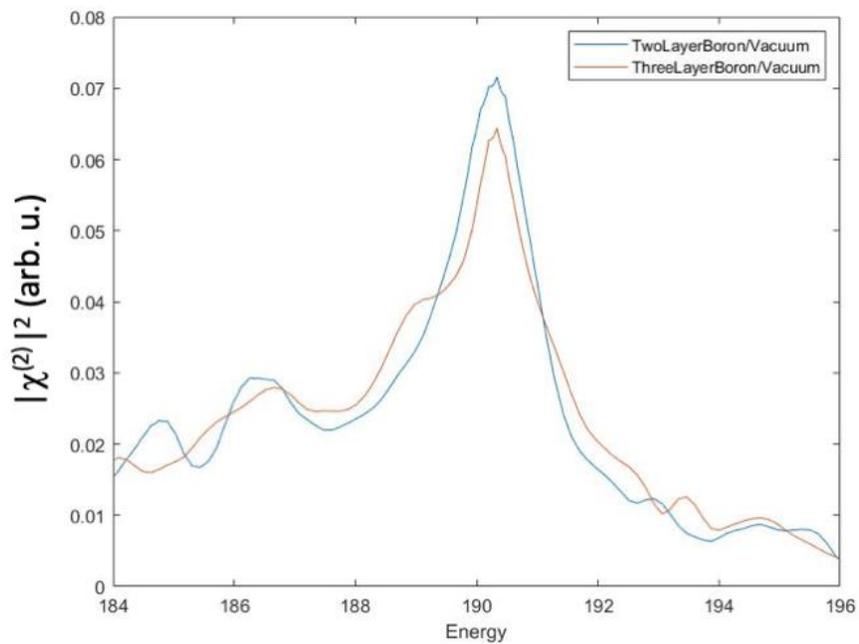

**Figure S8:** SXR-SHG spectra comparison between two layers (blue) and three layers (orange) of B/V interface.

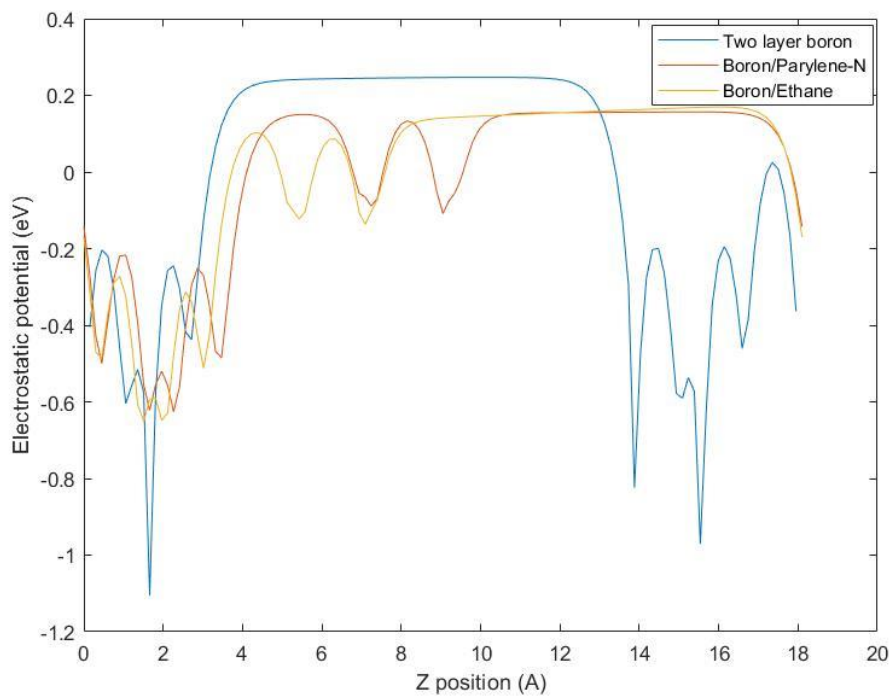

**Figure S9:** The calculated electrostatic potential at different z positions. The electrostatic potential as a function of position along the z axis (perpendicular to the slab) shows no dipole effect at the surface since the vacuum region (above ~3 Å) has a relatively constant potential. This is true for the B/V (blue), B/PN (orange) systems and B/E systems (yellow).

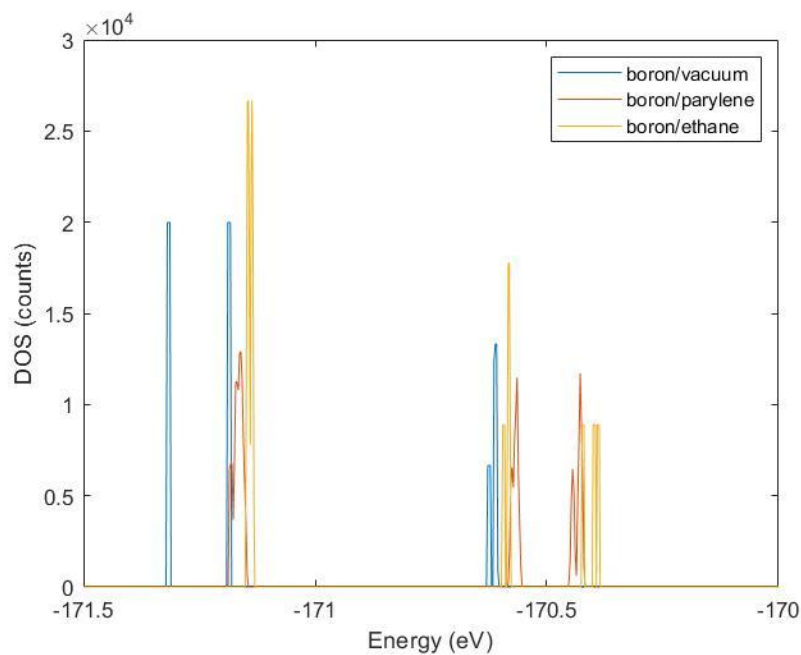

**Figure S10:** Calculated density of states for the interfaces studied here. The density of states of B/V (blue), B/PN(orange) and B/E (yellow) interfaces are shown.

**a)**

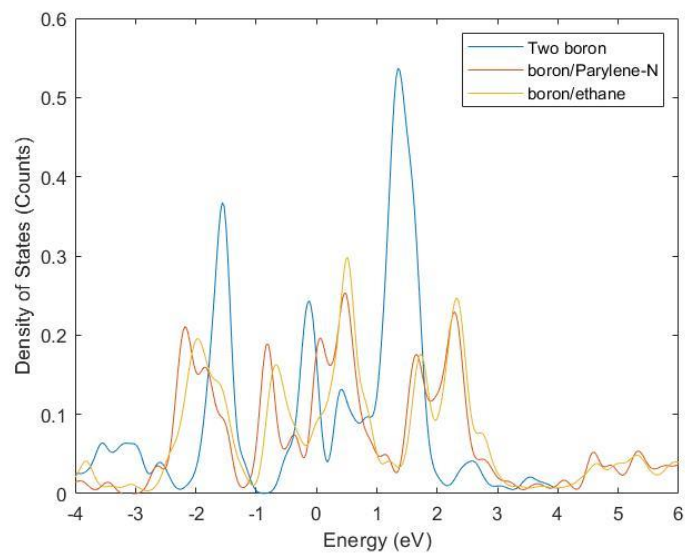

**b)**

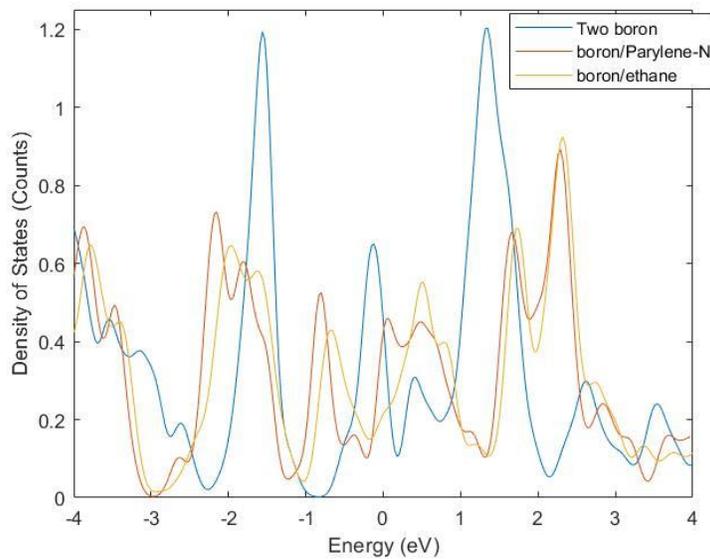

**Figure S11:** 2s and 2p contributions to the electronic density of states at the interface. The 2s orbital (a) and the 2p orbital (b) contributions to the electronic density of states at the interface are shown for the B/V (blue) B/PN (orange) and B/E (yellow) interfaces. There is more s-orbital character and more states near the Fermi level (0 eV) for the B/PN and B/E interface.